# Mechanics of Morphogenesis in Neural Development: *in vivo*, *in vitro*, and *in silico*

*Joseph Sutlive[1†], Hamed Seyyedhosseinzadeh[1,2 †], Zheng Ao[3 †], Haning Xiu[1], Kun Gou[4\*], Feng Guo[3\*], and Zi Chen[1\*]*

[1]Department of Surgery, Brigham and Women's Hospital/Harvard Medical School, Boston, MA 02115
[2]Department of Mechanical Engineering, University of South Carolina, Columbia, SC 29208
[3]Department of Intelligent Systems Engineering, Indiana University Bloomington, Bloomington, IN 47405
[4]Department of Mathematical, Physical, and Engineering Sciences, Texas A&M University–San Antonio, San Antonio, TX 78224, USA
[†] These authors contributed equally
[\*] Corresponding author. E-mail: zchen33@bwh.harvard.edu; fengguo@iu.edu; Kun.Gou@tamusa.edu

**Abstract**
Morphogenesis in the central nervous system has received intensive attention as elucidating fundamental mechanisms of morphogenesis will shed light on the physiology and pathophysiology of the developing central nervous system. Morphogenesis of the central nervous system is of a vast topic that includes important morphogenetic events such as neurulation and cortical folding. Here we review three types of methods used to improve our understanding of morphogenesis of the central nervous system: *in vivo* experiments, organoids (*in vitro*), and computational models (*in silico*). The *in vivo* experiments are used to explore cellular- and tissue-level mechanics and interpret them on the roles of neurulation morphogenesis. Recent advances in human brain organoids have provided new opportunities to study morphogenesis and neurogenesis to compensate for the limitations of *in vivo* experiments, as organoid models are able to recapitulate some critical neural morphogenetic processes during early human brain development. Due to the complexity and costs of *in vivo* and *in vitro* studies, a variety of computational models have been developed and used to explain the formation and morphogenesis of brain structures. We review and discuss the Pros and Cons of these methods and their usage in the studies on morphogenesis of the central nervous system. Notably, none of these methods alone is sufficient to unveil the biophysical mechanisms of morphogenesis, thus calling for the interdisciplinary approaches using a combination of these methods in order to test hypotheses and generate new insights on both normal and abnormal development of the central nervous system.

## 1. Introduction

Morphogenesis, the process that gives rise to distinctive shapes of functional organs such as the brain [1], heart [2], and lung [3], has fascinated researchers for a few centuries. Folding of cell sheets is a key strategy to shape tissues in embryogenesis, often accompanied by a constriction on one side of the sheet driven by molecular motors. While many genes regulating this constriction are identified, the biophysical principles that allow molecular and cellular forces to elicit cell sheet folding remain elusive. Examples of cell sheet folding include gastrulation in Drosophila [4] and neurulation in avian embryos [5], [6] that show remarkably similar folding mechanisms.

The morphogenetic process of the central nervous system has received special attention, as understanding neural tube and brain morphogenesis is paramount to finding ways to treat and prevent many neurological disorders that are rooted in early development. Our understanding of morphogenesis is dependent upon multidisciplinary approaches and studies from multiple perspectives. Broadly speaking, in order to develop invagination and folding, mechanical asymmetries and geometric frustration are typically needed in morphogenetic processes ranging from ventral furrow formation to neural tube formation, and from gut looping to cortical folding. Numerous studies have illustrated the molecular signaling pathways needed to initiate an invagination or folding. Nevertheless, tissue folding is, in essence, a biophysical process whereby mechanical forces drive the shaping of complex three-dimensional structures in organogenesis. Morphogenesis of the central nervous system is of a vast topic that includes important morphogenetic events such as neurulation and cortical folding. Here we review three types of methods used to improve our understanding of the mechanics of morphogenesis of the central nervous system: *in vivo* experiments, organoids, and computational models. Using these different methods, we can more comprehensively learn the biomechanical mechanisms of the development processes of the central nervous system, which will then shed light on the relationship between the structure and functionality.

## 2. In vivo experiments

There are several species typically used for embryonic morphogenesis research [7]–[11]. These models include species which lay eggs, such as birds and amphibians, and make incubation during specific timepoints easy to manage without the need for invasive surgeries. *In vivo* work often falls into one of four categories: neurogenesis, neurulation, organogenesis/early-stage morphogenesis, and cortical-folding/late-stage morphogenesis. Gastrulation, an early event leading to multiple germ layers in the embryo, has been shown to be dependent on localized active forces such as apical constriction through the accumulation of myosin II [12] and specific mechanical strain requirements (e.g., as shown in convergence/extension patterns) for invagination to occur [13]–[15]. This has been observed in multiple species including zebrafish[13], [16], Xenopus[14], [15], and Drosophila[16]. Zebrafish studies have additionally shown that the mechanical properties affecting gastrulation further affect the timing related to the onset of neurulation[13]. Additionally, it has been shown mechanical forces are critical in developing neurons as they play a critical role in axon extension[17].

Neurulation is another morphogenetic event in which a biomechanical perspective could aid in determining the driving factors of its onset and action. Primary neurulation is the morphogenetic process during which the neural tube (NT) forms from the two-dimensional neural plate (NP). The mechanisms of neurulation have been studied in amniotes such as chicks, mice, amphibians (e.g., frogs), and zebrafishes [11], for which conserved mechanisms have been identified. Convergent extension causes the NP to elongate and narrow [18], [19], and then the morphological change that follows is biphasic [20]: the first phase is when the medial hingepoint (MHP) forms, which bends the NP to form a V-shape neural groove (Fig. 1); this is followed by the elevation of neural folds (NFs) due to head mesenchyme expansion [21]. The second phase features the neural plate folding at the coupled dorso-lateral hingepoints, which brings the NFs closer and closer until fusion happens at the midline, thus forming the NT. In amniotes, NT closure of the cranial region starts at multiple locations and progresses through a zippering process to seal the NT. Oftentimes incomplete NT closure would occur, resulting in exencephaly and anencephaly [20].

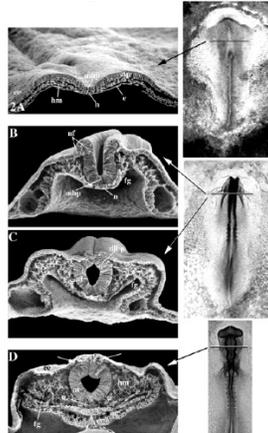

*Fig. 1. A schematic view (A-D) of the chick embryonic transverse sections employing the scanning electron microscopy during several time points of the primary neurulation at roughly the mesencephalon level. The right three subfigures are the orientation whole-mounts (3 insets to right). Modified from* [22].

Primary neurulation shares many similarities with gastrulation in terms of both the shape changes and mechanical mechanisms involved; in addition to neurulation being contingent upon the folding of cell sheets, it has been shown to be dependent on apical constricting forces, cell elongation, and cell migration[23]. In chick embryos, neurulation begins within the first day of incubation, as the NP gradually folds into the neural tube. The neurocoele (the cavity of the tube) terminates in openings called the rostral and caudal neuropores. At Hamilton and Hamburger (HH) stage 9 (~27 hr), the brain starts to expand differentially, and the major subdivisions of the brain take shape. Three vesicles (prosencephalon, mesencephalon, and rhombencephalon) and two bends (cervical flexure and cephalic flexure) arise in the cranial part of the neural tube (Fig. 1)[24]. By the HH stage 10 (~35 hr), the optic vesicles protrude from the prosencephalon. Further partitioning, bending, and bulging follow, during which the prosencephalon is partitioned into diencephalon and telencephalon, and the rhombencephalon evolves to form metencephalon and myelencephalon.

Several studies using Xenopus have been able to illustrate the mechanical system surrounding the events of neurulation and further determine the relationship of forces and mechanical properties to the success (or failure) of the development of the neural cord. It was recently shown that cells in early neural tissue are polar in a way that allows for coordinated actomyosin contractions[25]. This event, known as planar cell polarity[25], can be measured through a few different methods[26]. The main goal is to create a phenotype in which the structure of the tissue has obviously polarized (shape favoring one side or the other) the geometry. In this way, information about the cells can be obtained without the need for protein staining[26]. These types of contractions are often necessary to drive tissue in a consistent direction[25].

Several early-stage morphogenetic events, including heart looping and bending of the neural tube, have been shown to be driven by mechanical forces. Using chick embryos as a model, much progress has been made in determining the mechanics driving brain and heart morphogenesis. In the chick embryo, the embryo sits on top of the yolk between two membranes: the vitelline membrane sitting above the embryo, and the splanchnopleure membrane which lies below the embryo. For instance, it has been shown that the bending of the neural tube and twisting of the heart during Hamburger and Hamilton (HH) stages 10-14 is dependent upon the physical interaction of the vitelline membrane pushing against the embryo as it develops. This forces the embryo to break left-right asymmetry and twist [27]. To further illustrate the mechanical constraints on the system, it was shown that moving the position of the heart to the opposite side of the embryo resulted in twisting occurring in the opposite direction. Chick embryos have also been used in studies that determined the mechanical properties of the early brain are critical in neural crest migration.

Cortical folding is one of the key late-stage events of brain development. Substantial folding of the cerebral cortex in humans starts from the 5th fetal month and keeps on developing into the first year post-natal. This cortical folding enables the cortex surface area to reach 7 times larger than non-convoluted surface [28]. Many major gyri (outward folds) and sulci (inward folds) are consistently located and form well-known landmarks of the cortical surface, but in some regions the pattern of folds exhibits individual variability [29]. Major anomalies of cortical development include lissencephaly or agyria (absence of convolutions), microgyria (abnormally narrow folds), and pachygyria (broad gyri with thick cortex) [29]. These anomalies are generally in association with serious neurological disorders. More intricate disorders in folding have been shown to be associated with schizophrenia [30] and epilepsy [31]. The brains of many large mammals exhibit cortical folding (notably, the brains of mice and rats do not). The ferret is considered to be one of the smallest animals used in labs with its cerebral cortex [32], [33]; in addition, the folding process takes place postnatally, with the bulk of the process occurring during the first four weeks. Smart and McSherry [32], [33] described the geometric changes in the ferret brain from post-natal days 1–36. They describe the folds in terms of two gyral systems: (1) the spleniocruciate system (spleniocruciate and orbital gyrus), and (2) a dorsolateral system composed of the coronal and ectosylvian gyri. They postulate that gyri are formed where the cortex expands locally, both radially and tangentially, and sulci result simply from lack of radial growth.

Naturally, there have been many questions regarding cortical folding. One prevailing hypothesis was that the tension created by CNS axons pulling on the brain was a major contributor to cortical folding[28], [34], the so-called tension-based morphogenesis (TBM). It was later found that this might not necessarily be the case, as it has since been shown that the heterogeneous thicknesses amongst cell layers in the cortex are related to cortical folding[34], [35]. Additionally, it has been shown that the convoluted folding patterns are the result of mechanical stresses which in turn arise due to the growth of tissue[36]. Mechanical buckling is yet another proposed mechanism as shown by physical models and computational modelling that similar convolutions to human cortical folding (a real physical brain and a simulated brain shown in Fig. 2a-b) arise from faster expansion of an initially smooth layer bonded to a slowly enlarging area [37], [38]. These models corroborate with the findings that the relatively fast growth of the gray matter in the cortex results in folding and that the spatial-temporal variations in growth rate lead to the regional differences in folding [39]. Van Essen later proposed a "differential expansion sandwich plus (DES)" modification of the prior TBM hypothesis to interpret neural cortical enlargement and folding [40] (Fig. 2c). This revised model addresses the disadvantages of the original TBM model and include novel approaches that invite testing through in vivo experiments. For cerebellar cortex, he further proposed a cerebellar multilayer sandwich (CMS) model to include a variety of features such as the accordion-like folding in adult brain architecture (Fig. 2d).

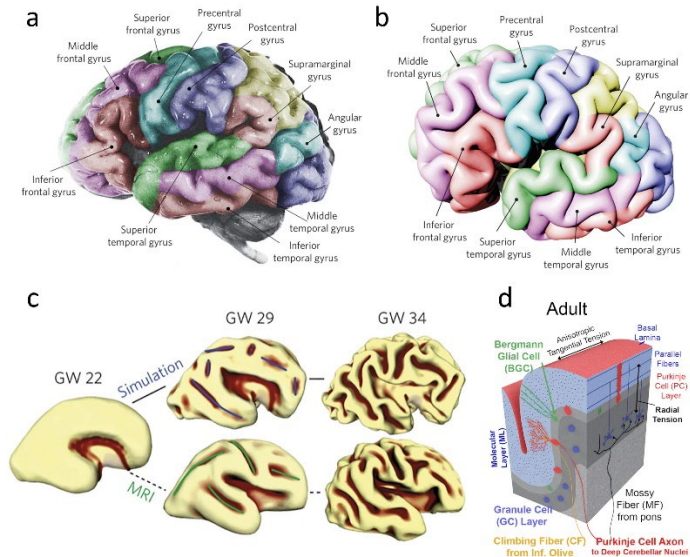

*Fig. 2. (**a**) A real brain and (**b**) a simulated brain illustrating all notable gyri. The foldings in (**b**) are driven by human cortical expansion. ((**a**) and (**b**) reprint permission needed from reference* [41]*); (**c**) Computational models with direct simulation (upper two subfigures) and simulation from MRI imaging (lower two subfigures) of cortical folding starting from GW 22 to GW 34. (**d**) Critical features of a cerebellar architecture of adult including Purkinje cells in red color, Bergmann radial glial cells in green color and granule cells in blue color. ((**c**) and (**d**) reprint permission needed from reference* [40]*).*

Although in vivo work has led to many interesting findings regarding morphogenesis across all stages, there are limitations to the current methods available. This usually arises from limitations in temporal and/or spatial resolution. Typically, a method with a high temporal resolution sacrifices spatial resolution and vice-versa. This provides the opportunity for interdisciplinary methodology to validate many of the findings *in vivo* and further explore the mechanical basis of morphogenesis. These methods tend to fall within two groupings: organoid (*in vitro*) and computational modeling (*in silico*) studies.

## 3. Organoids (in vitro)

Human brain organoids[42]–[44] are 3D brain-like *in vitro* cultures grown from human pluripotent stem cells (Fig. 3a). Brain organoids feature cellular components of the brain, cytoarchitecture, and even display certain neural activities similar to those in the brains. Therefore, they could recapitulate critical neural morphogenesis processes during early human brain development such as the formation of the ventricular zone (VZ), subventricular zone (SVZ), and human-specific outer subventricular zone (oSVZ) composed of outer radial glia cells (oRG), and cortical plates (CP)[42]. The unique mitosis and migration properties of oRG are believed to contribute to increased neural progenitor numbers and diversity in humans [45]. The mitotic oRG in human brain organoids, was shown to behave similarly to that of developing fetal brain and experience a specific division pattern: mitotic somal translocation (MST), where its soma translates to CP before cytokinesis [46]. Culturing brain organoids under a confined microchamber with real-time imaging has revealed that two opposing forces including the cytoskeletal contraction and nuclear expansions at the perimeter are the major contributing factors to organoid wrinkling, mimicking the physics of in vivo brain folding [47]. This MST process is thought to be an essential process in humans that mediates cortical layer expansion, and brain folding.

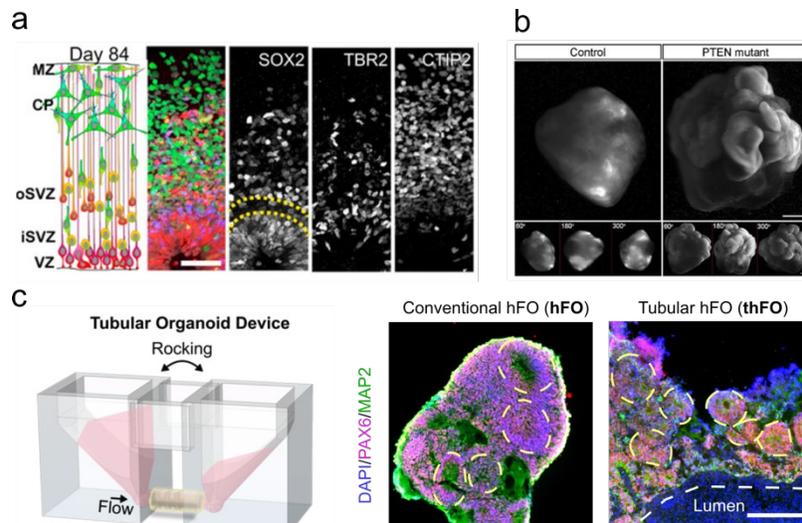

*Fig. 3. **(a)** Brain organoids recapitulate essential cytoarchitecture of the developing human brain (reprint permission needed from reference [48]); **(b)** Genetic engineering of human iPSC could be performed to study brain folding in organoids (reprint permission needed from reference [41]); **(c)** External force-induced through perfusion flow in tubular organoid cultures could alter neural morphogenesis (reprint permission needed from reference [49]).*

One major advantage of the human brain organoid models in neural morphogenesis study is that it provides an opportunity for developmental biologists to peek into early human brain development with unique human-specific genetics and disease-related perturbations, which would be largely inaccessible otherwise. Although some genetic functions and perturbations in brain development, cortical layer expansion, and brain folding are shared among human and animal models, some remain unique to humans such as regulation of GPR56 expression, as well as FGF2 signaling [50]. Ectopic modifications/perturbations could be easily interrogated in vitro using brain organoid models. For example, altering the mTOR pathway in oRG via shRNA was shown to modify oRG morphology and migration [51]. Additionally, the introduction of PTEN mutations by CRISPR/Cas9 could also induce expansion and folding in brain organoids (Fig. 3b) [41]. Interestingly, by replacing the human NOVA1 gene allele with that of Neanderthals, the brain organoids also show reduced neural progenitor cell (NPC) mitosis and altered morphology [52]. In addition to introduced perturbations, human brain organoids are capable of being grown from patient-derived induced pluripotent stem cells (iPSC) to examine disease-specific genetics and their effects on brain morphogenesis. For instance, brain organoids using cells from patients with severe microcephaly carrying truncating mutations in the CDK5RAP2 gene also show smaller neural epithelial tissue development and disrupted radial glial cell orientation [42]. Moreover, external factors, such as viral infection, environmental exposure to chemicals and drug abuse could alter brain development and morphogenesis as well. Applying external pathogenic agents to brain organoids could also reveal specific alternations they cause in human neural morphogenesis. For example, the ZIKA virus was shown to cause microcephaly and other severe brain defects during fetal development. Applying the ZIKA virus to brain organoids has revealed that they infect NPCs, including oRGs, resulting in decreased volumes in NPC and neuron layers[53]–[56]. Similarly, applying chemical agents such as alcohol, cannabis (Tetrahydrocannabinol, THC) and air pollutant (Particular Matter 2.5) have also been shown to induce developmental and morphological changes to brain organoids.

In addition to modeling the effects of human genetics and environmental factors on brain morphogenesis, brain organoids could also be employed to study how mechanical cues affect biological processes in brain morphogenesis in turn. During brain morphogenesis, the cortex gyrification is thought to guide neuron migration, and increase the surface area which leads to increased information processing ability in humans [37]. However, few studies have investigated how gyrification and interstitial pressure inside the neural tube in turn affect neural genesis. This could be largely attributed to the lack of appropriate models and engineering tools to interrogate such questions. Several studies have investigated how physical cues guide neural morphogenesis in brain organoids. By incorporating organoid culture into bioreactors, neural layer formation could be better guided with better nutrient perfusion [48]. By changing bioreactors from orbital mixing to vertical mixing, brain organoid morphology shows an inverted morphology with an NPC-outside, neuron-inside layering, as well as exhibiting ventral forebrain gene expression and cell identities [57]. By 3D scaffolding, brain organoids could also be guided into a neural tube configuration with internal lumen flows [49]. This configuration also showed increased NPC mitosis in the VZ/SVZ zone as well as an increased neuron layer in the cortical plate (Fig. 3c). Additionally, the constant perfusion through 3D scaffolds has also been shown to induce GABAergic neuron maturation and synapse formation [58].

Despite exciting discoveries that have been made using brain organoids as models to study neural morphogenesis, this field remains largely unexplored. Current organoids reflect partially the NPC mitosis and migration yet fail to reflect other physical forces in the brain, such as confining bone structures as well as intestinal pressure from embryonic cerebrospinal fluid (CSF). Additionally, brain vasculature also serves as important physical scaffolding and biological cues to guide brain morphogenesis and neuron migration/projection [59]. Furthermore, earlier developmental events, such as gastrulation and neurulation, although both have been partially modeled using organoid models, have yet been fully recapitulated as a single process in vitro [60], [61]. Further bioengineering development, as well as stem cell biology advances, are underway to tackle these problems.

## 4. Computational Methods
### 4.1 Types of models

At present, experimental measurement techniques and computational models are created to study and help understand the effects of interactive forces between cell and cell/tissue[62]. Experimental methods, for example, laser microsurgery [63] and traction force microscopy [64], were employed to measure cellular forces generated in morphogenesis and to unveil the mechanism of embryo and/or neurulation morphogenesis. However, computational methods, such as agent-based models[65]–[68] and vertex models [69]–[71], can be alternative techniques that provide information on mechanical processes of neuromorphogenesis when conventional experimental measurements are difficult to implement in vivo. We review and discuss computational methods that are used to measure/predict tissue forces in various morphogenesis.

a. *Vertex models*

Vertex models are proposed to generate results showing detailed cell-cell interactions through the driving of individual geometries making up "cells". Vertex models have another capability of capturing and visualizing topological transformation at the cellular level [72] with complex cell morphology. A 2D vertex dynamics model was built to predict formation of the epithelial tissue by Nagai and Honda [73]. They started using random Voronoi cell patterns to set up the initial state of cells and applying dynamics model to simulate the morphogenesis process. Their simulations illustrated the formation mechanism made cells to be uniform and shapes to become symmetric (homogeneous state). Their results had good agreements with the experiments observed by Honda et al [74], [75]. Later, Honda and his group extended the 2D model to a 3D version that can provide shapes of cell surfaces during morphogenesis [76]. This revised model can provide insight into the morphogenesis of viscoelastic tissues including glial cells and neurons. The model considered potential energy due to tension forces and energy dissipation from lateral forces (viscous drag) generated by cell interfaces. Their *in silico* model is capable of recognizing the shape transitions and rearrangements of cells in neurulation morphogenesis. The morphogenesis of the epithelial tissue was also studied using the computational model by Yu and Fernandez-Gonzalez [69].They combined biochemical dynamics with mechanical mechanism to investigate cellular mechanical forces in the dorsal closure process based on vertex models.

Vertex models have been developed to comprehensively study epithelial morphogenesis [71], [77]–[79], while they are still in the infancy of modeling specific mechanics of neurulation [80]. Nishimura et al. [81] applied the vertex model to clarify the responses of neuroepithelial cells activated by AJ-associated actomyosin. However, it is challenging to use vertex models to explore neurulation morphogenesis since the neurulation morphogenesis has complex cell and tissue movements [82], [83]. Studies of mechanics of morphogenesis in brain and nervous system are implemented using more generic measurements of extension and convergence[84], [85] combined with vertex-models (see an example in Fig. 4a). This combination can be helpful to validate and predict the intracellular forces and their effects on cell interfaces and tissue morphogenesis involved in neurulation.

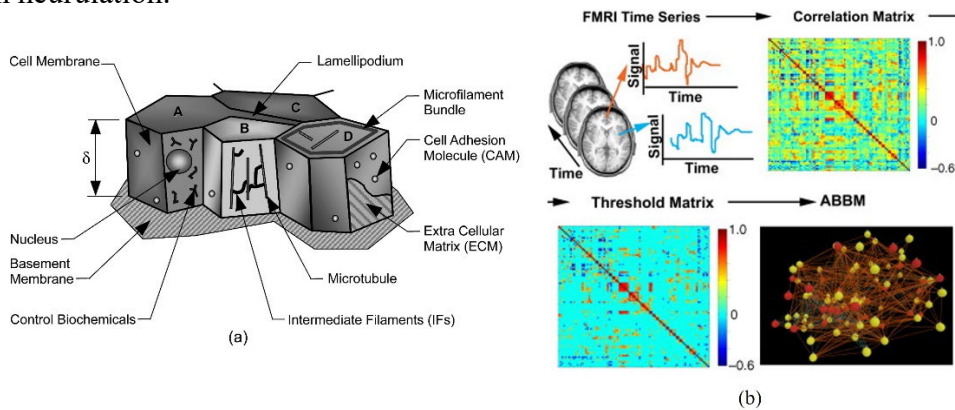

*Fig. 4. (a) Schematic of an epithelium with 3D vertex-model. (Reprint permission needed from reference [85]). (b) An agent-based brain model (ABBM) is created from a functional brain network in microscale level and is extended to mesoscale and macroscale. (Reprint permission needed from reference [86]).*

b. *Agent-based models*

Agent-based models are typically used to generate tissue-level biomechanical behaviors by modeling collections of biological cells. An agent-based model can be simplified to grid points ("sites") on a 2D/3D lattice with elastoviscous properties [87]. The model constrains that the cell movement can only have relative deformation between nearest neighboring grid points and the cell movement can then be explained by forces and energy in the system. Like vertex models, agent-based models are typically used to uncover new understandings of biomechanical mechanisms in epithelial and tissue morphogenesis, such as unveiling unicellular and multicellular biomechanical behaviors [65]–[67]. A modified agent-based model, called MecaGen, is capable of elucidating tissue transformations in a spatial-temporal domain [88]. It enables its user to explore cell dynamics in morphogenesis in both temporal and spatial domains.

Agent-based models of morphogenesis can be categorized into proliferation, migration, and differentiation using different agents to characterize various types of cells, e.g., stem cells and those differentiated ones [77]. Morphogenesis of the nervous system is associated with the movement of neural cells, which is one of three above-mentioned categories using agent-based models. For instance, several studies [89]–[91]qualitatively analyzed the advancement of the enteric nervous system associated with the movement of enteric neural crest cells (ENCCs). The ENCCs first had the same contribution to progeny as the original population, while they dominated the population after "invading" the gastrointestinal tract. Pennisi et. al [92] proposed an agent-based model to investigate the role of a basic function (blood–brain barrier) of the nervous system and then provided therapeutic strategies to these demyelinating diseases. However, these agent-based models alone are difficult to capture the evolution of complex tissue morphology. The brain and the central nervous system exhibit hierarchy in both the structural level and the functional level [86] , and therefore the computational models for the morphogenesis of the nervous system need to span cellular, tissue and organ levels. Joyce et. al [86] attempted to develop a brain-based mesoscale model employing the agent-based modeling approach with the rules developed in the microscale level but skillfully applied to the mesoscale and macroscale levels (Fig. 4b). Oscillatory, chaotic, and critical cell-cell (neuron-neuron) behaviors were observed while the model cannot recapitulate the anatomic and physiological structure of the brain.

Recently, an agent-based model of cell cloning has been developed to interpret the long-standing mystery of cerebellum folding [93]. In sharp contrast to cerebrum folding, the folding pattern in cerebellum is composed of curved surfaces with parallel grooves. It was not well understood what mechanism initiates and controlls the locations of those anchoring centers and how the cerebellar morphology is evolved biomechanically. Lejeune et al. [93]used an agent-based model to connect the cellular scale to the tissue scale, to identify the way that differential growth existing in the cerebellum results in geometric frustration for generation of those parallel grooves. Their study showed that the oriented granule cell division results in geometric instability and mechanical forces that caused the formation of the signature parallel folds in the cerebellum. This study generated new insight into the initiation of those anchoring centers in cerebellum and showcases the power of a multiscale biomechanical model for brain morphogenesis. Another example of a multiscale agent-based model is developed by Bauer et al[94] for simulating Cortical Layer Formation. It is composed of two stages of apoptosis and is suitable for the broad range of neuronal numbers in a variety of species. It is important to note that by incorporating apoptosis this model enables the variation of one layer's thickness without influencing other layers, which enriches the feasibility for evolutionary modification in layer architecture. It is interesting to note that modified gene regulatory dynamics can well recapitulate certain features in neurodevelopmental disorders, which further enables broader applications of this type of model for studying normal and abnormal cortical morphogenesis.

c. *Continuum models*

Continuum modeling considers tissues as continuum and are described by partial differential equations (PDE) or integro-differential equations[95], [96]. Most current continuum models in embryology are based on the simulation work of Alan Turing[97] which is reaction-diffusion (RD) modeling. Continuum models are also used to understand the biomechanical mechanisms within morphogenesis[98] The fundamental principles of continuum mechanics need to be satisfied including equilibria of linear and angular momentum, geometric compatibility, mass balance, and stress-strain constitutive relations[99]. Constitutive equations related to morphogenesis could be a combination of chemical, electric and mechanical equations.

Continuum models can also consider the brain tissue as a viscous liquid. The rationale of this modeling approach is due to the large time length during morphogenesis with the brain matter resembling a viscous fluid [100] and also due to the fact that cells in the early stages of morphogenetic processes have the eddy-like patterns and moves like what we see in fluid dynamics. Engstrom et al. [101] proposed a multi-phase model that include both elastic and fluid-like layers, with prescribed differential expansion between layers and radial constraints. Nevertheless, they did not take into consideration the cellular or tissue mechanics during the initiation of cerebellar folding probably due to technological limitations at that time. Based on experimental findings in the murine cerebellum that there was no obvious cellular pre-pattern at the initiation of folding and that the expansion of the outer layer expansion appears to be uniform and fluid-like, Lawton et al. [102]proposed that a multi-phase model can better recapitulate the evolution of configurations in cerebellar folding, especially at the initiation stage. In their model, the outer layer is fluidic and undergoes differential expansion while there are radial and circumferential constraints in the cerebellum.

Garcia et al. employs a mathematical modelling approach to interpret the development of chicken cerebral hemispheres subjected to a variety of experimental conditions[103]. Wang et al. develops a biomechanical model that incorporates cell division and migration as well as volumetric growth to illuminate the roles of biomechanical forces in the development of cortical folding [104]. Furthermore, a comprehensive review on computational modeling studies of cortical folding is provided by Darayi et al. [105]. These models, often accompanied by tissue cutting experiments [34], *in vivo* imaging [39], and/or physical model demonstration[38], [106], have enabled the testing of various hypotheses related to gaining mechanistic understanding of brain morphogenesis.

The similarities and distinctions between the cerebral cortex and the cerebellar cortex are worthy noting. They are both sheet-like structures yet there are significant structural differences. The cerebral cortex in humans has a total surface area of $1,843 \pm 196\ cm^2$ [107]and its thickness varies dramatically (as much as two-fold) across different regions with an average thickness of 2.6 mm [108]. In contrast, cerebellar cortex's thickness does not vary much and is typically less than 1 mm [109]. The major folds in the cerebellar cortex are accordion-like [110], [111]. In both the cerebral cortex and the cerebellum, radial glia spans the entire structure [112]. Moreover, in some species outer radial glial cells have been evolved to enable the development of folded cortexes [113], for instance, studies on how Bergmann glia have been inducted in regulating the cerebellar cortex folding have been reviewed by Leung and Li [114].

## 4.2 Applications of models
a. *2D models for epithelial invagination*

The neural tube bears dorsolateral hinge points in some regions by which folding can occur due to cell proliferation and migration. During epithelial invagination, the cells in the neural tube can slide or push against each other due to various cellular forces including apical constriction forces. The mechanism of such bending remains incompletely understood and in terms of both morphological changes and mechanical forces involved such invagination is similar to ventral furrow formation in Drosophila gastrulation. Brezavscek et al [115]studied a 2D model for the epithelial invagination formed by mechanical forces of identical cells. This model treated all cells to be of the same type and ignored the difference of cells in the folding region. The purpose is to test whether arrangement of cells functions produce epithelial invagination. The authors employed the ventral furrow formation in *Drosophila melanogaster*, the fruit fly, as an example to test the model's effectiveness. The model is based on the cross-section of a tube-like epithelium like the early *Drosophila* embryo and assumed that such cross-section bears all the needed components for the physical approach using the Hamiltonian surface energy [116], [117]. Each cell is simplified to be of the shape of a quadrilateral, and N such cells are arranged one by one to form a ring-like shape with the ring surrounding the yolk. The area of each cell and the area of the yolk are constrained to be constant during the invagination process [118]. The outer boundary of the ring is surrounded by a shell with changeable stiffness to simulate the vitelline membrane [118]. For each quadrilateral representing the cell, their boundaries can be categorized into three types consisting of the basal side (towards the center of the ring) with the length $L_b$, the apical side (towards the outer boundary of the ring) with the length $L_a$, and the two lateral sides (shared with neighboring cells) with a total length $L_l$. The tensions over the three sides are denoted by $\Gamma_b$, $\Gamma_a$, and $\Gamma_l$, respectively (Fig. 5).

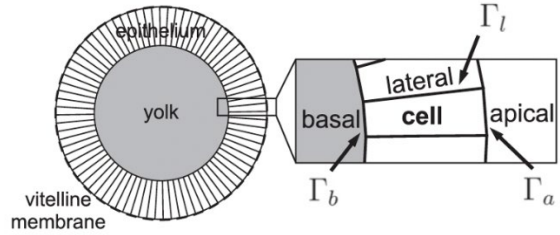

*Fig. 5. A schematic graph for the epithelium composed of cells in quadrilateral shapes surrounding the yolk. The vitelline membrane is over the outer boundary of the epithelium. The enlarged quadrilateral shows the basal, apical, and lateral sides. The tensions over the three sides are denoted by $\Gamma_b$, $\Gamma_a$, and $\Gamma_l$, respectively. (The figure is from [115] with the reproduction permission obtained.)*

Then the total energy $W_r$ from the cell-formed ring is provided by

$$W_r = \sum_{i=1}^{N} \Gamma_b L_b^i + \Gamma_a L_a^i + \Gamma_l L_l^i, \quad (1)$$

where the superscript "$i$" over $L$ means the $i_{th}$ cell in the ring. Such equation can be nondimensionalized for convenience of analysis and computation using some reference values. The shell surrounding the ring is elastic and will be expanded once the ring size increases [87], producing a pressure over the ring satisfying

$$P(r) = P_0 \left( \exp\left( \frac{r - r_v}{r_v} \right) - 1 \right), \text{ for } r > r_v. \quad (2)$$

Here $P_0 = 10^4 P_{int}$ with $P_{int}$ being the pressure in the yolk before the ring buckling, $r_v$ is the undeformed radius of the shell under no pressing force from the cell ring, and $r$ is the distance of any point to the ring center. When $r < r_v$, $P(r) = 0$ because the shell membrane cannot shrink and thus creates no pressure over the ring. The shell membrane provides another energy part $W_m$ to the system as

$$W_m = \sum_{i=1}^{N} P(r_i) \Delta A_i. \quad (3)$$

Here $\Delta A_i$ is the area of each cell outside of the undeformed shell vitelline membrane, and $r_i$ is the radius of the centroid of such area. The total energy of the system is $W = W_r + W_m$. We can fix $\Gamma_l$ and vary $\Gamma_b$ and $\Gamma_a$ to minimize the energy $W$. The results show that such model using energy minimization also generates epithelial invagination. The extent of invagination can be controlled by the parameters $\Gamma_b$ and $\Gamma_a$. The simulation results match the lab outcome using *Drosophila*.
[115]

b. *Bilayer shell models to describe cortical folding*

When the brain develops, it increases its volume and surface area under the process of cortical folding. The origination of such process is complicated involving factors such as radial constraint [119], differential growth [120], and axon maturation [121]. The time scale for the cortical folding is several months consisting of the primary, secondary, and tertiary stages. Almost all human brains show identical developmental processes during the primary stage. However, different brains show different developmental processes during the secondary and tertiary stages. During the maturation, axon also grows to form a network throughout the brain [122]. While the axon-tension hypothesis had been popular for a long period of time, it has been challenged when new evidence emerged[34]. Many researchers then revisited and revived the idea that differential growth plays a key role in cortical folding[123]–[125], i.e., the cortical folding is driven by the differential growth rates between the surface layer and the inner core. Indeed, recent studies demonstrated that differential growth among the inner and outer brain layers may play a pivotal role for brain gyrification during the later secondary and tertiary stages [37] along with other factors such as axonal fiber locations and growth heterogeneity. The outer cortex layer can grow faster than the inner white-matter core, generating compressive forces over the white matter to form cortical folding with concave sulci and convex gyri. It was also found that cortical folding creates a pulling force on the axon to induce the inner white matter growth. Budday et al. [124]. built a mechanical model that incorporates both differential growth and axonal tension, two seemingly competing hypotheses to interpret the formation of morphological abnormalities in cortical folding. Their model applied nonlinear field theory of mechanics and was carefully calibrated using magnetic resonance images of the developing brains of preterm neonates at 27, 29 and 32 weeks of gestation. According to the model predictions, morphological abnormalities are derived from the misbalance in cortical growth rates and deviations in thickness. They further showed through simulations that those abnormalities were consistent with the characteristics of lissencephaly and polymicrogyria pathologies.

More recently, Chavoshnejad et al. [126] set up finite-element models to study how the axonal fibers with differential growth of the two layers are involved in the cortical folding. Particularly, they studied how the interaction of the differential growth, tension over the fiber axon, and fiber distribution and density generate different cortical folding patterns. Their results showed that different growth rates in the two layers is the main factor for the cortical folding, and that the cortical gyrification can exert a pulling force over the area with dense stiff fiber bundles towards the direction of gyri.

c. *Perspectives of modeling organoids*

The human brain is very complex, thus making deep studies for brain functionality and effective treatment of brain diseases challenging. Instead, the brain organoid obtained from the human induced pluripotent stem cells greatly facilitate deep studies on the brain structure and functionality [127], [128]. The organoid is formed through the self-organizing of the brain cells and the bio-engineered brain scaffold. Such organoid is not a perfect replica of the human brain, but faithfully recapitulates the main components of it. The organoid can provide a powerful tool to understand the origin of neurological diseases and the corresponding treatments [129]. The 3D organoid model is advantageous over the 2D culture model in more intricately expressing the brain spatial structure, cell-cell and cell-matrix network [130]. Through the brain organoid, we can take live imaging of the brain activity in the cellular level for more convenient examination. The artificial organoid also helps us understand complicated interactions inside the brain such as neuron-glia and circuitry. It can be employed to study cognitive and psychiatric pathologies with polygenetic interactions, unknown genetic defects, or epigenetic alterations [131]. The pathologies studied using the brain organoids include, but are not limited to, microcephaly, macrocephaly, autism, Rett syndrome, Miller-Dieker syndrome, and Sandhoff disease. As the gene editing techniques develop, the brain organoid can be used to study how gene mutation or repairs can help treat the neurological disorders.

The brain organoid also bears limitations. The brain organoid is a separate organ-like structure and does not bear the surrounding supportive tissue and axial patterning signals as a natural brain. This limits the organoid's capability to develop into the in-vivo brain structure. The organoid also lacks the vascularization to provide oxygen or nutrients into the organoid tissue, a possible reason for low progenitor amount and for induction of necrosis in the organoid center [132]. To overcome such drawback, microfluidic device can be used to transport nutrient into the organoid. Transplantation of the brain organoids into animals can facilitate growth of vascularization into the organoid for faster maturation of the organoid[133]. The brain organoid takes a long time to grow into a mature one suitable for research, which significantly reduces research efficiency and increases the culture cost [134]. Techniques for speeding up the maturation are therefore very necessary. Moreover, no two brain organoids are identical due to variable growth factors, and new techniques should be created to reduce these variations for growing more similar brain organoids and making such research results more reproducible.

**4.3 Validation, limitations, and future work**

Morphogenesis events are the combination of mechanical and chemical rules[88] at different length scales; i.e., they have multiscale multiphysics processes controlling final biological structures and behaviors (Fig. 7).

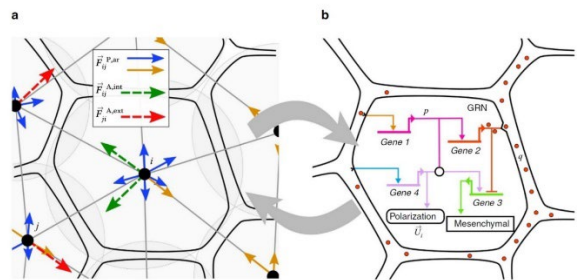

*Fig. 7. A schematic view of the coupling of the cell's biomechanical properties to its biochemical activity. (a) Cell shapes and the related mechanical forces to maintain the cell shapes and dynamics, (b) the biochemical model derived from a gene regulatory network driven by chemical kinetics and reaction-diffusion equations*[88].

Current models use assumptions and certain techniques to simplify the simulations and complexity of the underlying physics at some length scales, so they cannot completely replace laboratory experiments and find all the biophysical mechanisms[135]. Furthermore, although there are many tailored models for specific morphogenetic events in certain species, none of these systems can offer a fully integrated framework to establish the relationships among molecular signaling, genetic regulation, force regulation, cell migration, growth, and pattern formation [88]. Despite heavy modelling efforts, models alone cannot provide convincing causal relationship between a specific mechanism and a specific observed result. Models can show that special mechanisms can help explaining a specific observed result, but modeling alone is insufficient to provide the evidence that such mechanisms are solely responsible for that result. The limitations of current models might be related to some other underlying mechanisms, perhaps yet unknown. Models can, nevertheless, disprove certain mechanisms with high confidence. In the event that there are only limited sets of mechanisms for an observed result (e.g., cortical folding), they are useful in disproving the hypothesis (or hypotheses) that a given set of driving forces suffice in generating an observed result. Examples include some early two-dimensional (2D) models of neurulation [136], [137]

Tissues and organs are three-dimensional (3D). Therefore, the 3D models provide important insights into the roles of cell mechanics in morphogenesis [76]. Typically, 3D models share the same mechanical concepts as their 2D counterparts, although there are scenarios that 3D models contain information that cannot be well characterized in 2D models [138], [139]. For example, the 3D model for invagination is found to be more robust to perturbations (on the ratio between apical constriction and apicobasal elongation) than the 2D counterpart [140].

Although hybrid models are complicated, they will work to recapitulate tissue dynamics more faithfully. This branch, typically referred to as multiscale multiphysics modeling within the mathematics and engineering communities, integrates different tissue components, relations within these components, and other information from multiple scales in one model[141]. Hybrid models are usually set up on the conservative laws of mass, momentum, and energy using continuum equations for mechanical deformation of viscous medium and/or fluid dynamics coupled with averaged discrete and stochastic methods like Cellular Automata and Pott model considering growth kinetics, rheology, etc.[142], [143]. The hybrid models can also use quantum mechanics' first principal methods like density functional theory (DFT) to address the gaps through different scales instead of stochastic methods. Moreover, hybrid models based on continuum equations coupled with quantum methods allow the fine-tunning and validation of the cell-related parameters obtained from microscale measurements[144]. This kind of hybrid model is promising for simulation results as close as an experiment.

Finally, we do not elaborate on neurite growth and morphogenesis due to the length limit although such processes are crucial in neural development[145]. Neurites include axons that transmits nervous signals and dendrites that receive such signals. During neural development, the neurites sprout out, migrate, grow to identify the pathways to form interconnections with other neurons. A variety of genetic, biochemical and biomechanical factors together drive this process and mathematical models play pivotal roles in testing various hypotheses regarding the growth and morphogenesis of neurites. An excellent review on this topic has been provided by Oliveri and Goriely [145], which focuses on the mechanical aspects of neuronal growth and morphogenesis.

## 5. Summary and Conclusion

The *in vivo* experiments, *in vitro* organoid models, and *in silico* computational modeling all provide effective means to study morphogenesis of the nervous system. The *in vivo* work mainly employ experiments on live animals to study neurogenesis, neurulation, organogenesis (early-stage morphogenesis), and cortical-folding (late-stage morphogenesis). Gastrulation has been observed across different species. Primary neurulations are very similar to gastrulation in the aspect of folding of cell sheets but also depend on apical constricting forces, cell elongation, and cell migration to complete the task[23]. Several early-stage morphogenetic events, including bending and twisting of the neural tube, have been shown to be driven by the interaction of mechanical forces.

Due to the limitation of *in vivo* experiments, iPSCs derived human brain organoids have been employed to study morphogenesis and neurogenesis. The artificial organoids are capable of recapitulating critical early neural morphogenesis processes for areas such as the ventricular zone, subventricular zone, human-specific outer subventricular zone, and cortical plates [42]. The human brain organoid models can greatly assist developmental biologists to study early human brain development under the genetic effect and disease-related perturbations on morphogenesis, and how mechanical cues affect the developmental processes. Brain organoids represent a novel and rapidly developing area. This technique sheds light on a deeper exploration of brain development, its functionality, and corresponding neurological disease studies in human systems. Many challenging issues arise in culturing the brain organoids to closely simulate the *in vivo* counterpart and developmental trajectory.

The *in vivo* and *in vitro* models take time to grow and may not be sufficient for identifying the causal relationships among different parts of the early developing brain. To address this issue, *in silico* models taking into effect the physical principles can largely facilitate understanding the formation of the spatial and temporal brain structures. Vertex models, for instance, have been used to study cell-cell interactions considering each cell as a driving force to the collective system and are capable of both capturing and visualizing topological transformation at the cellular level [72]. As an example, Brezavscek et al. [62] studied a 2D model for the epithelial invagination formed by mechanical forces of identical cells using the energy minimization approach to create invagination successfully. In comparison, agent-based models usually generate tissue-level biomechanical behaviors by considering aggregates of cells. This type of model can be simplified to grid points on lattices with elastoviscous properties[87]. For instance, the work in [126] set up nonlinear finite-element models to study how the axonal fibers with differential growth of the inner and outer brain layers are involved in cortical folding. Their models involved interaction of the differential growth, tension over the fiber axon, and fiber distribution and density, and successfully obtained different cortical folding results.

Nonetheless, none of the three methods can independently unveil secrets of morphogenesis of the central nervous system. The in vivo experiments can provide us with the most direct results of this process but with limited data. The organoid models can provide more data but with less accuracy. The computational models simplified the brain structure and help us explore the internal relations of the main brain structures in the developmental processes. Each one bears its advantages and disadvantages. Only when combined, can they complement each other to generate new insight on these complicated yet important biological processes. We are still in the early stage of such exploration. Interdisciplinary collaboration from areas studying molecular biology, cellular biology, tissue engineering, biochemistry, biophysics, and biomathematics can extraordinarily advance the development of this area.

**Acknowledgement**

This work was supported in part by the support from the National Science Foundation (Award #2025434) and Fund to Sustain Research Excellence from Brigham Research Institute. Kun Gou is grateful to the support from the College Grant Program of College of Arts and Sciences at Texas A&M University-San Antonio.

**References:**
[1] R. Gordon and G. W. Brodland, "The cytoskeletal mechanics of brain morphogenesis," *Cell Biophysics 1987 11:1*, vol. 11, no. 1, pp. 177–238, Dec. 1987, doi: 10.1007/BF02797122.

[2] J. M. Schleich, T. Abdulla, R. Summers, and L. Houyel, "An overview of cardiac morphogenesis," *Archives of Cardiovascular Diseases*, vol. 106, no. 11, pp. 612–623, Nov. 2013, doi: 10.1016/J.ACVD.2013.07.001.

[3] J. C. Schittny, "Development of the lung," *Cell and Tissue Research*, vol. 367, no. 3, p. 427, Mar. 2017, doi: 10.1007/S00441-016-2545-0.

[4] M. Leptin, "Gastrulation in Drosophila: The logic and the cellular mechanisms," *EMBO Journal*, vol. 18, no. 12. John Wiley & Sons, Ltd, pp. 3187–3192, Jun. 15, 1999. doi: 10.1093/emboj/18.12.3187.

[5] N. Kinoshita, N. Sasai, K. Misaki, and S. Yonemura, "Apical Accumulation of Rho in the Neural Plate Is Important for Neural Plate Cell Shape Change and Neural Tube Formation," *https://doi.org/10.1091/mbc.e07-12-1286*, vol. 19, no. 5, pp. 2289–2299, Mar. 2008, doi: 10.1091/MBC.E07-12-1286.

[6] D. S. Vijayraghavan and L. A. Davidson, "Mechanics of neurulation: From classical to current perspectives on the physical mechanics that shape, fold, and form the neural tube," *Birth Defects Research*, vol. 109, no. 2, pp. 153–168, Jan. 2017, doi: 10.1002/BDRA.23557.

[7] M. G. Davey and C. Tickle, "The chicken as a model for embryonic development," *Cytogenet Genome Res*, vol. 117, no. 1–4, pp. 231–239, Jul. 2007, doi: 10.1159/000103184.

[8] D. P. Kiehart, J. M. Crawford, A. Aristotelous, S. Venakides, and G. S. Edwards, "Cell Sheet Morphogenesis: Dorsal Closure in Drosophila melanogaster as a Model System," *https://doi.org/10.1146/annurev-cellbio-111315-125357*, vol. 33, pp. 169–202, Oct. 2017, doi: 10.1146/ANNUREV-CELLBIO-111315-125357.

[9] J. Liu and D. Y. R. Stainier, "Zebrafish in the study of early cardiac development," *Circulation Research*, vol. 110, no. 6, pp. 870–874, Mar. 2012, doi: 10.1161/CIRCRESAHA.111.246504.

[10] A. S. Warkman and P. A. Krieg, "Xenopus as a model system for vertebrate heart development," *Seminars in Cell and Developmental Biology*, vol. 18, no. 1, pp. 46–53, 2007, doi: 10.1016/J.SEMCDB.2006.11.010.

[11] J. M. Werner, M. Y. Negesse, D. L. Brooks, A. R. Caldwell, J. M. Johnson, and R. M. Brewster, "Hallmarks of primary neurulation are conserved in the zebrafish forebrain," *Communications Biology 2021 4:1*, vol. 4, no. 1, pp. 1–16, Jan. 2021, doi: 10.1038/s42003-021-01655-8.

[12] N. H. Granholm and J. R. Baker, "Cytoplasmic microtubules and the mechanism of avian gastrulation," *Dev Biol*, vol. 23, no. 4, pp. 563–584, 1970, doi: 10.1016/0012-1606(70)90141-7.

[13] D. Bhattacharya, J. Zhong, S. Tavakoli, A. Kabla, and P. Matsudaira, "Strain maps characterize the symmetry of convergence and extension patterns during zebrafish gastrulation," *Sci Rep*, vol. 11, no. 1, Dec. 2021, doi: 10.1038/S41598-021-98233-Z.

[14] R. Keller and M. Danilchik, "Regional expression, pattern and timing of convergence and extension during gastrulation of Xenopus laevis," 1988.

[15] R. Keller, J. Shih, and C. Domingo, "The patterning and functioning of protrusive activity during convergence and extension of the Xenopus organiser," *Journal of Neuroscience*, vol. 13, no. 6, pp. 81–91, 1993.

[16]  B. He, K. Doubrovinski, O. Polyakov, and E. Wieschaus, "Apical constriction drives tissue-scale hydrodynamic flow to mediate cell elongation," *Nature*, vol. 508, no. 7496, pp. 392–396, Mar. 2014, doi: 10.1038/nature13070.

[17]  M. A. Breau, I. Bonnet, J. Stoufflet, J. Xie, S. De Castro, and S. Schneider-Maunoury, "Extrinsic mechanical forces mediate retrograde axon extension in a developing neuronal circuit," *NatCo*, vol. 8, no. 1, p. 282, Dec. 2017, doi: 10.1038/S41467-017-00283-3.

[18]  A. J. Copp, N. D. E. Greene, and J. N. Murdoch, "Dishevelled: linking convergent extension with neural tube closure," *Trends in Neurosciences*, vol. 26, no. 9, pp. 453–455, Sep. 2003, doi: 10.1016/S0166-2236(03)00212-1.

[19]  R. Keller, "Shaping the vertebrate body plan by polarized embryonic cell movements," *Science (1979)*, vol. 298, no. 5600, pp. 1950–1954, Dec. 2002, doi: 10.1126/SCIENCE.1079478.

[20]  A. J. Copp, N. D. E. Greene, and J. N. Murdoch, "The genetic basis of mammalian neurulation," *Nature Reviews Genetics 2003 4:10*, vol. 4, no. 10, pp. 784–793, Oct. 2003, doi: 10.1038/nrg1181.

[21]  G. M. Morriss-Kay, "Growth and development of pattern in the cranial neural epithelium of rat embryos during neurulation," *Development*, vol. 65, no. Supplement, pp. 225–241, Oct. 1981, doi: 10.1242/DEV.65.SUPPLEMENT.225.

[22]  J. F. Colas and G. C. Schoenwolf, "Towards a cellular and molecular understanding of neurulation," *Dev Dyn*, vol. 221, no. 2, pp. 117–145, 2001, doi: 10.1002/DVDY.1144.

[23]  Y. Inoue *et al.*, "Mechanical roles of apical constriction, cell elongation, and cell migration during neural tube formation in Xenopus," *Biomech Model Mechanobiol*, vol. 15, no. 6, pp. 1733–1746, Dec. 2016, doi: 10.1007/S10237-016-0794-1.

[24]  D. Purves, G. Augustine, D. Fitzpatrick, A.-S. Lamantia, J. McNamara, and S. Williams, *Neuroscience*, 3rd ed. Sunderland, MA: Sinauer, 2004.

[25]  M. Matsuda and S. Y. Sokol, "Xenopus neural tube closure: A vertebrate model linking planar cell polarity to actomyosin contractions," *Curr Top Dev Biol*, vol. 145, pp. 41–60, Jan. 2021, doi: 10.1016/BS.CTDB.2021.04.001.

[26]  J. Olofsson and J. D. Axelrod, "Methods for studying planar cell polarity," *Methods*, vol. 68, no. 1, p. 97, Jun. 2014, doi: 10.1016/J.YMETH.2014.03.017.

[27]  Z. Chen, Q. Guo, E. Dai, N. Forsch, and L. A. Taber, "How the embryonic chick brain twists," *Journal of the Royal Society Interface*, vol. 13, no. 124, pp. 1–8, 2016, doi: 10.1098/rsif.2016.0395.

[28]  D. C. van Essen, "A tension-based theory of morphogenesis and compact wiring in the central nervous system," *Nature 1997 385:6614*, vol. 385, no. 6614, pp. 313–318, Jan. 1997, doi: 10.1038/385313a0.

[29]  W. Welker, "Why Does Cerebral Cortex Fissure and Fold?," E. Jones and A. Peters, Eds. New York: Plenum, 1990, pp. 3–136. doi: 10.1007/978-1-4615-3824-0_1.

[30]  P. C. Sallet *et al.*, "Reduced cortical folding in schizophrenia: an MRI morphometric study," *Am J Psychiatry*, vol. 160, no. 9, pp. 1606–1613, Sep. 2003, doi: 10.1176/APPI.AJP.160.9.1606.

[31]  K. M. Jacobs, M. Mogensen, E. Warren, and D. A. Prince, "Experimental Microgyri Disrupt the Barrel Field Pattern in Rat Somatosensory Cortex," *Cerebral Cortex*, vol. 9, no. 7, pp. 733–744, Oct. 1999, doi: 10.1093/CERCOR/9.7.733.

[32] I. H. M. Smart and G. M. Mcsherry, "Gyrus formation in the cerebral cortex in the ferret. I. Description of the external changes.," *Journal of Anatomy*, vol. 146, p. 141, 1986, Accessed: Jul. 03, 2022. [Online]. Available: /pmc/articles/PMC1166530/?report=abstract

[33] I. H. Smart and G. M. McSherry, "Gyrus formation in the cerebral cortex of the ferret. II. Description of the internal histological changes.," *Journal of Anatomy*, vol. 147, p. 27, Aug. 1986, Accessed: Jul. 03, 2022. [Online]. Available: /pmc/articles/PMC1261544/?report=abstract

[34] G. Xu, A. K. Knutsen, K. Dikranian, C. D. Kroenke, P. v. Bayly, and L. A. Taber, "Axons pull on the brain, but tension does not drive cortical folding," *J Biomech Eng*, vol. 132, no. 7, Jul. 2010, doi: 10.1115/1.4001683.

[35] L. da Costa Campos, R. Hornung, G. Gompper, J. Elgeti, and S. Caspers, "The role of thickness inhomogeneities in hierarchical cortical folding," *Neuroimage*, vol. 231, May 2021, doi: 10.1016/J.NEUROIMAGE.2021.117779.

[36] M. J. Razavi, T. Zhang, X. Li, T. Liu, and X. Wang, "Role of mechanical factors in cortical folding development," *Phys Rev E Stat Nonlin Soft Matter Phys*, vol. 92, no. 3, Sep. 2015, doi: 10.1103/PHYSREVE.92.032701.

[37] T. Tallinen, J. Y. Chung, J. S. Biggins, and L. Mahadevan, "Gyrification from constrained cortical expansion," *Proc Natl Acad Sci U S A*, vol. 111, no. 35, pp. 12667–12672, Sep. 2014, doi: 10.1073/PNAS.1406015111/-/DCSUPPLEMENTAL.

[38] T. Tallinen, J. Y. Chung, F. Rousseau, N. Girard, J. Lefèvre, and L. Mahadevan, "On the growth and form of cortical convolutions," *Nature Physics 2016 12:6*, vol. 12, no. 6, pp. 588–593, Feb. 2016, doi: 10.1038/nphys3632.

[39] K. E. Garcia *et al.*, "Dynamic patterns of cortical expansion during folding of the preterm human brain," *Proc Natl Acad Sci U S A*, vol. 115, no. 12, pp. 3156–3161, Mar. 2018, doi: 10.1073/PNAS.1715451115/SUPPL_FILE/PNAS.201715451SI.PDF.

[40] D. C. van Essen, "A 2020 view of tension-based cortical morphogenesis," *Proc Natl Acad Sci U S A*, vol. 117, no. 52, pp. 32868–32879, Dec. 2020, doi: 10.1073/PNAS.2016830117/SUPPL_FILE/PNAS.2016830117.SAPP.PDF.

[41] Y. Li *et al.*, "Induction of Expansion and Folding in Human Cerebral Organoids," *Cell Stem Cell*, vol. 20, no. 3, pp. 385-396.e3, Mar. 2017, doi: 10.1016/J.STEM.2016.11.017.

[42] M. A. Lancaster *et al.*, "Cerebral organoids model human brain development and microcephaly," *Nature 2013 501:7467*, vol. 501, no. 7467, pp. 373–379, Aug. 2013, doi: 10.1038/nature12517.

[43] N. D. Amin and S. P. Paşca, "Building Models of Brain Disorders with Three-Dimensional Organoids," *Neuron*, vol. 100, no. 2, pp. 389–405, Oct. 2018, doi: 10.1016/J.NEURON.2018.10.007.

[44] C. A. Trujillo *et al.*, "Complex Oscillatory Waves Emerging from Cortical Organoids Model Early Human Brain Network Development," *Cell Stem Cell*, vol. 25, no. 4, pp. 558-569.e7, Oct. 2019, doi: 10.1016/J.STEM.2019.08.002.

[45] N. Matsumoto, S. Tanaka, T. Horiike, Y. Shinmyo, and H. Kawasaki, "A discrete subtype of neural progenitor crucial for cortical folding in the gyrencephalic mammalian brain," *Elife*, vol. 9, Apr. 2020, doi: 10.7554/ELIFE.54873.

[46] B. E. L. Ostrem, J. H. Lui, C. C. Gertz, and A. R. Kriegstein, "Control of outer radial glial stem cell mitosis in the human brain," *Cell Rep*, vol. 8, no. 3, pp. 656–664, Aug. 2014, doi: 10.1016/J.CELREP.2014.06.058.

[47] E. Karzbrun, A. Kshirsagar, S. R. Cohen, J. H. Hanna, and O. Reiner, "Human brain organoids on a chip reveal the physics of folding," *Nature Physics 2018 14:5*, vol. 14, no. 5, pp. 515–522, Feb. 2018, doi: 10.1038/s41567-018-0046-7.

[48] X. Qian *et al.*, "Brain-Region-Specific Organoids Using Mini-bioreactors for Modeling ZIKV Exposure," *Cell*, vol. 165, no. 5, pp. 1238–1254, May 2016, doi: 10.1016/J.CELL.2016.04.032.

[49] Z. Ao *et al.*, "Tubular human brain organoids to model microglia-mediated neuroinflammation," *Lab on a Chip*, vol. 21, no. 14, pp. 2751–2762, Jul. 2021, doi: 10.1039/D1LC00030F.

[50] B. G. Rash, S. Tomasi, H. D. Lim, C. Y. Suh, and F. M. Vaccarino, "Cortical Gyrification Induced by Fibroblast Growth Factor 2 in the Mouse Brain," *The Journal of Neuroscience*, vol. 33, no. 26, p. 10802, Jun. 2013, doi: 10.1523/JNEUROSCI.3621-12.2013.

[51] M. G. Andrews, L. Subramanian, and A. R. Kriegstein, "Mtor signaling regulates the morphology and migration of outer radial glia in developing human cortex," *Elife*, vol. 9, pp. 1–21, Sep. 2020, doi: 10.7554/ELIFE.58737.

[52] C. A. Trujillo *et al.*, "Reintroduction of the archaic variant of NOVA1 in cortical organoids alters neurodevelopment," *Science*, vol. 371, no. 6530, Feb. 2021, doi: 10.1126/SCIENCE.AAX2537.

[53] X. Qian, H. N. Nguyen, F. Jacob, H. Song, and G. L. Ming, "Using brain organoids to understand Zika virus-induced microcephaly," *Development*, vol. 144, no. 6, p. 952, Mar. 2017, doi: 10.1242/DEV.140707.

[54] J. Dang *et al.*, "Zika Virus Depletes Neural Progenitors in Human Cerebral Organoids through Activation of the Innate Immune Receptor TLR3," *Cell Stem Cell*, vol. 19, no. 2, pp. 258–265, Aug. 2016, doi: 10.1016/J.STEM.2016.04.014.

[55] F. R. Cugola *et al.*, "The Brazilian Zika virus strain causes birth defects in experimental models," *Nature 2016 534:7606*, vol. 534, no. 7606, pp. 267–271, May 2016, doi: 10.1038/nature18296.

[56] Q. Liang *et al.*, "Zika Virus NS4A and NS4B Proteins Deregulate Akt-mTOR Signaling in Human Fetal Neural Stem Cells to Inhibit Neurogenesis and Induce Autophagy," *Cell Stem Cell*, vol. 19, no. 5, pp. 663–671, Nov. 2016, doi: 10.1016/J.STEM.2016.07.019.

[57] D. N. A. Suong *et al.*, "Induction of inverted morphology in brain organoids by vertical-mixing bioreactors," *Communications Biology 2021 4:1*, vol. 4, no. 1, pp. 1–13, Oct. 2021, doi: 10.1038/s42003-021-02719-5.

[58] S. Song *et al.*, "Understanding immune-driven brain aging by human brain organoid microphysiological analysis platform," *bioRxiv*, p. 2022.01.19.476989, Jan. 2022, doi: 10.1101/2022.01.19.476989.

[59] M. Segarra, B. C. Kirchmaier, and A. Acker-Palmer, "A vascular perspective on neuronal migration," *Mech Dev*, vol. 138 Pt 1, pp. 17–25, Nov. 2015, doi: 10.1016/J.MOD.2015.07.004.

[60]  J.-H. Lee *et al.*, "Human spinal cord organoids exhibiting neural tube morphogenesis for a quantifiable drug screening system of neural tube defects," *bioRxiv*, p. 2020.12.02.409177, Dec. 2020, doi: 10.1101/2020.12.02.409177.

[61]  S. C. van den Brink and A. van Oudenaarden, "3D gastruloids: a novel frontier in stem cell-based in vitro modeling of mammalian gastrulation," *Trends Cell Biol*, vol. 31, no. 9, pp. 747–759, Sep. 2021, doi: 10.1016/J.TCB.2021.06.007.

[62]  K. Sugimura, P. F. Lenne, and F. Graner, "Measuring forces and stresses in situ in living tissues," *Development (Cambridge)*, vol. 143, no. 2, pp. 186–196, Jan. 2016, doi: 10.1242/dev.119776.

[63]  M. S. Hutson *et al.*, "Forces for morphogenesis investigated with laser microsurgery and quantitative modeling," *Science (1979)*, vol. 300, no. 5616, pp. 145–149, 2003, doi: 10.1126/science.1079552.

[64]  W. J. Polacheck and C. S. Chen, "Measuring cell-generated forces : a guide to the available tools," vol. 13, no. 5, pp. 415–423, 2017, doi: 10.1038/nmeth.3834.Measuring.

[65]  D. Drasdo and S. Höhme, "A single-cell-based model of tumor growth in vitro: Monolayers and spheroids," *Physical Biology*, vol. 2, no. 3, pp. 133–147, Sep. 2005, doi: 10.1088/1478-3975/2/3/001.

[66]  P. van Liedekerke *et al.*, "A quantitative high-resolution computational mechanics cell model for growing and regenerating tissues," *Biomechanics and Modeling in Mechanobiology*, vol. 19, no. 1, pp. 189–220, Feb. 2020, doi: 10.1007/s10237-019-01204-7.

[67]  P. van Liedekerke, M. M. Palm, N. Jagiella, and D. Drasdo, "Simulating tissue mechanics with agent-based models: concepts, perspectives and some novel results," *Computational Particle Mechanics*, vol. 2, no. 4, pp. 401–444, Dec. 2015, doi: 10.1007/s40571-015-0082-3.

[68]  M. H. Swat, G. L. Thomas, J. M. Belmonte, A. Shirinifard, D. Hmeljak, and J. A. Glazier, "Multi-Scale Modeling of Tissues Using CompuCell3D," in *Methods in Cell Biology*, vol. 110, Academic Press Inc., 2012, pp. 325–366. doi: 10.1016/B978-0-12-388403-9.00013-8.

[69]  J. C. Yu and R. Fernandez-Gonzalez, "Quantitative modelling of epithelial morphogenesis: integrating cell mechanics and molecular dynamics," *Seminars in Cell and Developmental Biology*, vol. 67. Elsevier Ltd, pp. 153–160, Jul. 01, 2017. doi: 10.1016/j.semcdb.2016.07.030.

[70]  S. Alt, P. Ganguly, and G. Salbreux, "Vertex models: from cell mechanics to tissue morphogenesis," *Philosophical Transactions of the Royal Society B: Biological Sciences*, vol. 372, no. 1720, p. 20150520, May 2017, doi: 10.1098/rstb.2015.0520.

[71]  P. Spahn and R. Reuter, "A Vertex Model of Drosophila Ventral Furrow Formation," *PLoS ONE*, vol. 8, no. 9, p. e75051, Sep. 2013, doi: 10.1371/journal.pone.0075051.

[72]  D. Pastor-Escuredo and J. C. del Álamo, "How Computation Is Helping Unravel the Dynamics of Morphogenesis," *Frontiers in Physics*, vol. 8, p. 31, Feb. 2020, doi: 10.3389/fphy.2020.00031.

[73]  T. Nagai and H. Honda, "A dynamic cell model for the formation of epithelial tissues," *https://doi.org/10.1080/13642810108205772*, vol. 81, no. 7, pp. 699–719, 2009, doi: 10.1080/13642810108205772.

[74] H. Honda, R. Kodama, T. Takeuchi, H. Yamanaka, K. Watanabe, and G. Eguchi, "Cell behaviour in a polygonal cell sheet," *Development*, vol. 83, no. Supplement, pp. 313–327, Nov. 1984, doi: 10.1242/DEV.83.SUPPLEMENT.313.

[75] H. Honda, Y. Ogita, S. Higuchi, and K. Kani, "Cell movements in a living mammalian tissue: Long-term observation of individual cells in wounded corneal endothelia of cats," *Journal of Morphology*, vol. 174, no. 1, pp. 25–39, Oct. 1982, doi: 10.1002/JMOR.1051740104.

[76] H. Honda, M. Tanemura, and T. Nagai, "A three-dimensional vertex dynamics cell model of space-filling polyhedra simulating cell behavior in a cell aggregate," *J Theor Biol*, vol. 226, no. 4, pp. 439–453, Feb. 2004, doi: 10.1016/J.JTBI.2003.10.001.

[77] R. Farhadifar, J. C. Röper, B. Aigouy, S. Eaton, and F. Jülicher, "The influence of cell mechanics, cell-cell interactions, and proliferation on epithelial packing," *Curr Biol*, vol. 17, no. 24, pp. 2095–2104, Dec. 2007, doi: 10.1016/J.CUB.2007.11.049.

[78] P. Sahlin and H. Jönsson, "A Modeling Study on How Cell Division Affects Properties of Epithelial Tissues Under Isotropic Growth," *PLOS ONE*, vol. 5, no. 7, p. e11750, 2010, doi: 10.1371/JOURNAL.PONE.0011750.

[79] O. Polyakov, B. He, M. Swan, J. W. Shaevitz, M. Kaschube, and E. Wieschaus, "Passive mechanical forces control cell-shape change during drosophila ventral furrow formation," *Biophysical Journal*, vol. 107, no. 4, pp. 998–1010, Aug. 2014, doi: 10.1016/j.bpj.2014.07.013.

[80] D. S. Vijayraghavan and L. A. Davidson, "Mechanics of neurulation: From classical to current perspectives on the physical mechanics that shape, fold, and form the neural tube," *Birth Defects Research*, vol. 109, no. 2, pp. 153–168, Jan. 2017, doi: 10.1002/BDRA.23557.

[81] T. Nishimura, H. Honda, and M. Takeichi, "Planar Cell Polarity Links Axes of Spatial Dynamics in Neural-Tube Closure," *Cell*, vol. 149, no. 5, pp. 1084–1097, May 2012, doi: 10.1016/J.CELL.2012.04.021.

[82] A. G. Fletcher, M. Osterfield, R. E. Baker, and S. Y. Shvartsman, "Vertex Models of Epithelial Morphogenesis," *Biophysical Journal*, vol. 106, no. 11, p. 2291, Jun. 2014, doi: 10.1016/J.BPJ.2013.11.4498.

[83] P. Guerrero *et al.*, "Neuronal differentiation influences progenitor arrangement in the vertebrate neuroepithelium," *Development (Cambridge)*, vol. 146, no. 23, Dec. 2019, doi: 10.1242/DEV.176297/VIDEO-1.

[84] M. Zajac, G. L. Jones, and J. A. Glazier, "Simulating convergent extension by way of anisotropic differential adhesion," *J Theor Biol*, vol. 222, no. 2, pp. 247–259, May 2003, doi: 10.1016/S0022-5193(03)00033-X.

[85] G. W. Brodland and J. H. Veldhuis, "Lamellipodium-driven tissue reshaping: a parametric study," *Comput Methods Biomech Biomed Engin*, vol. 9, no. 1, pp. 17–23, 2006, doi: 10.1080/10255840600554703.

[86] K. E. Joyce, P. J. Laurienti, and S. Hayasaka, "Complexity in a brain-inspired agent-based model," *Neural Netw*, vol. 33, pp. 275–290, Sep. 2012, doi: 10.1016/J.NEUNET.2012.05.012.

[87]  C. M. Glen, M. L. Kemp, and E. O. Voit, "Agent-based modeling of morphogenetic systems: Advantages and challenges," *PLoS Computational Biology*, vol. 15, no. 3, p. e1006577, Mar. 2019, doi: 10.1371/journal.pcbi.1006577.

[88]  J. Delile, M. Herrmann, N. Peyriéras, and R. Doursat, "A cell-based computational model of early embryogenesis coupling mechanical behaviour and gene regulation," *Nature Communications 2017 8:1*, vol. 8, no. 1, pp. 1–10, Jan. 2017, doi: 10.1038/ncomms13929.

[89]  J. I. Lake and R. O. Heuckeroth, "Enteric nervous system development: migration, differentiation, and disease," *Am J Physiol Gastrointest Liver Physiol*, vol. 305, no. 1, Jul. 2013, doi: 10.1152/AJPGI.00452.2012.

[90]  N. Nagy and A. M. Goldstein, "Enteric nervous system development: A crest cell's journey from neural tube to colon," *Semin Cell Dev Biol*, vol. 66, pp. 94–106, Jun. 2017, doi: 10.1016/J.SEMCDB.2017.01.006.

[91]  D. Zhang, I. M. Brinas, B. J. Binder, K. A. Landman, and D. F. Newgreen, "Neural crest regionalisation for enteric nervous system formation: implications for Hirschsprung's disease and stem cell therapy," *Dev Biol*, vol. 339, no. 2, pp. 280–294, Mar. 2010, doi: 10.1016/J.YDBIO.2009.12.014.

[92]  M. Pennisi, A. M. Rajput, L. Toldo, and F. Pappalardo, "Agent based modeling of Treg-Teff cross regulation in relapsing-remitting multiple sclerosis," *BMC Bioinformatics*, vol. 14, no. Suppl 16, p. S9, Oct. 2013, doi: 10.1186/1471-2105-14-S16-S9.

[93]  E. Lejeune, B. Dortdivanlioglu, E. Kuhl, and C. Linder, "Understanding the mechanical link between oriented cell division and cerebellar morphogenesis," *Soft Matter*, vol. 15, no. 10, pp. 2204–2215, Mar. 2019, doi: 10.1039/C8SM02231C.

[94]  R. Bauer, G. J. Clowry, and M. Kaiser, "Creative Destruction: A Basic Computational Model of Cortical Layer Formation," *Cerebral Cortex (New York, NY)*, vol. 31, no. 7, p. 3237, Jul. 2021, doi: 10.1093/CERCOR/BHAB003.

[95]  V. Cristini, X. Li, J. S. Lowengrub, and S. M. Wise, "Nonlinear simulations of solid tumor growth using a mixture model: invasion and branching," *Journal of Mathematical Biology 2008 58:4*, vol. 58, no. 4, pp. 723–763, Sep. 2008, doi: 10.1007/S00285-008-0215-X.

[96]  J. S. Lowengrub *et al.*, "Nonlinear modelling of cancer: bridging the gap between cells and tumours," *Nonlinearity*, vol. 23, no. 1, p. R1, 2010, doi: 10.1088/0951-7715/23/1/R01.

[97]  A. M. Turing, "The Chemical Basis of Morphogenesis," *Philosophical Transactions of the Royal Society of London. Series B, Biological Sciences*, vol. 237, no. 641, pp. 37–72, 1952, Accessed: Jun. 30, 2022. [Online]. Available: http://www.jstor.org/about/terms.html.

[98]  Y. Shi, J. Yao, G. Xu, and L. A. Taber, "Bending of the Looping Heart: Differential Growth Revisited," *Journal of Biomechanical Engineering*, vol. 136, no. 8, p. 0810021, 2014, doi: 10.1115/1.4026645.

[99]  W. M. Lai, D. Rubin, and E. Krempl, *Introduction to Continuum Mechanics*. Elsevier Inc., 2010. doi: 10.1016/B978-0-7506-8560-3.X0001-1.

[100]  G. Forgacs, R. A. Foty, Y. Shafrir, and M. S. Steinberg, "Viscoelastic properties of living embryonic tissues: a quantitative study.," *Biophysical Journal*, vol. 74, no. 5, p. 2227, 1998, doi: 10.1016/S0006-3495(98)77932-9.

[101] T. A. Engstrom, T. Zhang, A. K. Lawton, A. L. Joyner, and J. M. Schwarz, "Buckling without Bending: A New Paradigm in Morphogenesis," *Physical Review X*, vol. 8, no. 4, p. 041053, Dec. 2018, doi: 10.1103/PHYSREVX.8.041053/FIGURES/7/MEDIUM.

[102] A. K. Lawton *et al.*, "Cerebellar folding is initiated by mechanical constraints on a fluid-like layer without a cellular pre-pattern," *Elife*, vol. 8, Apr. 2019, doi: 10.7554/ELIFE.45019.

[103] K. E. Garcia, W. G. Stewart, M. Gabriela Espinosa, J. P. Gleghorn, and L. A. Taber, "Molecular and mechanical signals determine morphogenesis of the cerebral hemispheres in the chicken embryo," *Development (Cambridge)*, vol. 146, no. 20, Oct. 2019, doi: 10.1242/DEV.174318/VIDEO-3.

[104] S. Wang, K. Saito Id, H. Kawasaki Id, and M. A. Holland Id, "Orchestrated neuronal migration and cortical folding: A computational and experimental study," *PLOS Computational Biology*, vol. 18, no. 6, p. e1010190, Jun. 2022, doi: 10.1371/JOURNAL.PCBI.1010190.

[105] M. Darayi *et al.*, "Computational models of cortical folding: A review of common approaches," *Journal of Biomechanics*, vol. 139, p. 110851, Jun. 2022, doi: 10.1016/J.JBIOMECH.2021.110851.

[106] P. v. Bayly, L. A. Taber, and C. D. Kroenke, "Mechanical forces in cerebral cortical folding: A review of measurements and models," *Journal of the Mechanical Behavior of Biomedical Materials*, vol. 29, pp. 568–581, Jan. 2014, doi: 10.1016/J.JMBBM.2013.02.018.

[107] C. J. Donahue, M. F. Glasser, T. M. Preuss, J. K. Rilling, and D. C. van Essen, "Quantitative assessment of prefrontal cortex in humans relative to nonhuman primates," *Proc Natl Acad Sci U S A*, vol. 115, no. 22, pp. E5183–E5192, May 2018, doi: 10.1073/PNAS.1721653115/-/DCSUPPLEMENTAL.

[108] M. F. Glasser *et al.*, "The Human Connectome Project's Neuroimaging Approach," *Nat Neurosci*, vol. 19, no. 9, p. 1175, Aug. 2016, doi: 10.1038/NN.4361.

[109] "The Human Brain Atlas at Michigan State University." https://brains.anatomy.msu.edu/brains/human/index.html (accessed Jul. 03, 2022).

[110] D. C. van Essen, "Surface-based atlases of cerebellar cortex in the human, macaque, and mouse," *Ann N Y Acad Sci*, vol. 978, pp. 468–479, 2002, doi: 10.1111/J.1749-6632.2002.TB07588.X.

[111] S. C. Sereno, G. G. Scott, B. Yao, E. J. Thaden, and P. J. O'Donnell, "Emotion word processing: does mood make a difference?," *Frontiers in Psychology*, vol. 6, p. 1191, Aug. 2015, doi: 10.3389/FPSYG.2015.01191.

[112] M. Götz, E. Hartfuss, and P. Malatesta, "Radial glial cells as neuronal precursors: a new perspective on the correlation of morphology and lineage restriction in the developing cerebral cortex of mice," *Brain Res Bull*, vol. 57, no. 6, pp. 777–788, 2002, doi: 10.1016/S0361-9230(01)00777-8.

[113] I. Reillo, C. de Juan Romero, M. Á. García-Cabezas, and V. Borrell, "A Role for Intermediate Radial Glia in the Tangential Expansion of the Mammalian Cerebral Cortex," *Cerebral Cortex*, vol. 21, no. 7, pp. 1674–1694, Jul. 2011, doi: 10.1093/CERCOR/BHQ238.

[114] A. W. Leung and J. Y. H. Li, "The Molecular Pathway Regulating Bergmann Glia and Folia Generation in the Cerebellum," *The Cerebellum 2017 17:1*, vol. 17, no. 1, pp. 42–48, Dec. 2017, doi: 10.1007/S12311-017-0904-3.

[115] A. Hocevar Brezavšček, M. Rauzi, M. Leptin, and P. Ziherl, "A model of epithelial invagination driven by collective mechanics of identical cells," *Biophys J*, vol. 103, no. 5, pp. 1069–1077, Sep. 2012, doi: 10.1016/J.BPJ.2012.07.018.

[116] J. J. Muñoz, K. Barrett, and M. Miodownik, "A deformation gradient decomposition method for the analysis of the mechanics of morphogenesis," *Journal of Biomechanics*, vol. 40, no. 6, pp. 1372–1380, 2007, doi: 10.1016/J.JBIOMECH.2006.05.006.

[117] V. Conte, J. J. Mũnoz, B. Baum, and M. Miodownik, "Robust mechanisms of ventral furrow invagination require the combination of cellular shape changes," *Physical Biology*, vol. 6, no. 1, p. 016010, Apr. 2009, doi: 10.1088/1478-3975/6/1/016010.

[118] P. A. Pouille and E. Farge, "Hydrodynamic simulation of multicellular embryo invagination," *Phys Biol*, vol. 5, no. 1, Apr. 2008, doi: 10.1088/1478-3975/5/1/015005.

[119] P. Rakic, "A small step for the cell, a giant leap for mankind: a hypothesis of neocortical expansion during evolution," *Trends Neurosci*, vol. 18, no. 9, pp. 383–388, 1995, doi: 10.1016/0166-2236(95)93934-P.

[120] J. H. E. Cartwright, "Labyrinthine Turing pattern formation in the cerebral cortex," *J Theor Biol*, vol. 217, no. 1, pp. 97–103, 2002, doi: 10.1006/JTBI.2002.3012.

[121] D. C. van Essen, "A tension-based theory of morphogenesis and compact wiring in the central nervous system," *Nature 1997 385:6614*, vol. 385, no. 6614, pp. 313–318, Jan. 1997, doi: 10.1038/385313a0.

[122] D. H. Roossien, P. Lamoureux, and K. E. Miller, "Cytoplasmic dynein pushes the cytoskeletal meshwork forward during axonal elongation," *J Cell Sci*, vol. 127, no. Pt 16, pp. 3593–3602, 2014, doi: 10.1242/JCS.152611.

[123] D. P. Richman, R. M. Stewart, J. W. Hutchinson, and V. S. Caviness, "Mechanical model of brain convolutional development," *Science*, vol. 189, no. 4196, pp. 18–21, 1975, doi: 10.1126/SCIENCE.1135626.

[124] S. Budday, C. Raybaud, and E. Kuhl, "A mechanical model predicts morphological abnormalities in the developing human brain," *Scientific Reports 2014 4:1*, vol. 4, no. 1, pp. 1–7, Jul. 2014, doi: 10.1038/srep05644.

[125] S. Budday, P. Steinmann, and E. Kuhl, "Physical biology of human brain development," *Frontiers in Cellular Neuroscience*, vol. 9, no. JULY, p. 257, Jul. 2015, doi: 10.3389/FNCEL.2015.00257/BIBTEX.

[126] P. Chavoshnejad *et al.*, "Role of axonal fibers in the cortical folding patterns: A tale of variability and regularity," *Brain Multiphysics*, vol. 2, Jan. 2021, doi: 10.1016/J.BRAIN.2021.100029.

[127] S. P. Medvedev, A. I. Shevchenko, and S. M. Zakian, "Induced Pluripotent Stem Cells: Problems and Advantages when Applying them in Regenerative Medicine," *Acta Naturae*, vol. 2, no. 2, p. 18, Jun. 2010, doi: 10.32607/20758251-2010-2-2-18-27.

[128] G. Quadrato, J. Brown, and P. Arlotta, "The promises and challenges of human brain organoids as models of neuropsychiatric disease," *Nat Med*, vol. 22, no. 11, pp. 1220–1228, Nov. 2016, doi: 10.1038/NM.4214.

[129] Y. Shi, H. Inoue, J. C. Wu, and S. Yamanaka, "Induced pluripotent stem cell technology: a decade of progress," *Nat Rev Drug Discov*, vol. 16, no. 2, pp. 115–130, Feb. 2017, doi: 10.1038/NRD.2016.245.

[130] H. Wang, "Modeling Neurological Diseases With Human Brain Organoids," *Frontiers in Synaptic Neuroscience*, vol. 10, p. 15, Jun. 2018, doi: 10.3389/FNSYN.2018.00015/BIBTEX.

[131] S. L. Forsberg, M. Ilieva, and T. Maria Michel, "Epigenetics and cerebral organoids: promising directions in autism spectrum disorders," *Translational Psychiatry 2017 8:1*, vol. 8, no. 1, pp. 1–11, Jan. 2018, doi: 10.1038/s41398-017-0062-x.

[132] S. L. Giandomenico and M. A. Lancaster, "Probing human brain evolution and development in organoids," *Curr Opin Cell Biol*, vol. 44, pp. 36–43, Feb. 2017, doi: 10.1016/J.CEB.2017.01.001.

[133] A. A. Mansour *et al.*, "An in vivo model of functional and vascularized human brain organoids," *Nat Biotechnol*, vol. 36, no. 5, pp. 432–441, Jun. 2018, doi: 10.1038/NBT.4127.

[134] X. Qian, H. Song, and G. L. Ming, "Brain organoids: advances, applications and challenges," *Development*, vol. 146, no. 8, Apr. 2019, doi: 10.1242/DEV.166074.

[135] G. W. Brodland, "How computational models can help unlock biological systems," *Semin Cell Dev Biol*, vol. 47–48, pp. 62–73, Dec. 2015, doi: 10.1016/J.SEMCDB.2015.07.001.

[136] G. W. Brodland and D. A. Clausi, "Simulation of Morphogenetic Shape Changes Using the Finite Element Method," *Biomechanics of Active Movement and Division of Cells*, pp. 425–430, 1994, doi: 10.1007/978-3-642-78975-5_13.

[137] D. A. Clausi and G. W. Brodland, "Mechanical evaluation of theories of neurulation using computer simulations," *Development*, vol. 118, no. 3, pp. 1013–1023, Jul. 1993, doi: 10.1242/DEV.118.3.1013.

[138] R. Allena, A. S. Mouronval, and D. Aubry, "Simulation of multiple morphogenetic movements in the Drosophila embryo by a single 3D finite element model.," *J Mech Behav Biomed Mater*, vol. 3, no. 4, pp. 313–324, Jan. 2010, doi: 10.1016/J.JMBBM.2010.01.001.

[139] X. Du, M. Osterfield, and S. Y. Shvartsman, "Computational analysis of three-dimensional epithelial morphogenesis using vertex models," *Phys Biol*, vol. 11, no. 6, 2014, doi: 10.1088/1478-3975/11/6/066007.

[140] V. Conte, J. J. Muñoz, and M. Miodownik, "A 3D finite element model of ventral furrow invagination in the Drosophila melanogaster embryo," *Journal of the Mechanical Behavior of Biomedical Materials*, vol. 1, no. 2, pp. 188–198, Apr. 2008, doi: 10.1016/J.JMBBM.2007.10.002.

[141] L. A. D'Alessandro, S. Hoehme, A. Henney, D. Drasdo, and U. Klingmüller, "Unraveling liver complexity from molecular to organ level: Challenges and perspectives," *Progress in Biophysics and Molecular Biology*, vol. 117, no. 1, pp. 78–86, Jan. 2015, doi: 10.1016/J.PBIOMOLBIO.2014.11.005.

[142] H. Byrne and D. Drasdo, "Individual-based and continuum models of growing cell populations: a comparison," *J Math Biol*, vol. 58, no. 4–5, pp. 657–687, Apr. 2009, doi: 10.1007/S00285-008-0212-0.

[143] P. Ghysels, G. Samaey, B. Tijskens, P. van Liedekerke, H. Ramon, and D. Roose, "Multi-scale simulation of plant tissue deformation using a model for individual cell mechanics," *Phys Biol*, vol. 6, no. 1, 2009, doi: 10.1088/1478-3975/6/1/016009.

[144] F. Milde, M. Bergdorf, and P. Koumoutsakos, "A Hybrid Model for Three-Dimensional Simulations of Sprouting Angiogenesis," *Biophysical Journal*, vol. 95, no. 7, p. 3146, Oct. 2008, doi: 10.1529/BIOPHYSJ.107.124511.

[145] H. Oliveri and A. Goriely, "Mathematical models of neuronal growth," *Biomechanics and Modeling in Mechanobiology 2022 21:1*, vol. 21, no. 1, pp. 89–118, Jan. 2022, doi: 10.1007/S10237-021-01539-0.